\documentclass[aps,prd,secnumarabic,amssymb, amsmath,nobibnotes,nofootinbib,11pt]{revtex4}
\usepackage{enumitem}
\usepackage{amsfonts}
\usepackage{amssymb}
\usepackage{graphicx}
\usepackage{array}
\usepackage{amsbsy}
\usepackage{bbold}
\usepackage{appendix}
\usepackage{mathrsfs}
\usepackage{hyperref}
\usepackage{amsmath}
\usepackage{physics}
\usepackage{amsthm}
\usepackage{cancel}
\usepackage{yfonts}
\usepackage{inputenc}
\usepackage{chngcntr}

\newcommand{\be}{\begin{equation}}
\newcommand{\bal}{\begin{align}}
\newcommand{\eal}{\end{align}}
\newcommand{\ee}{\end{equation}}
\newcommand{\bea}{\begin{eqnarray}}
\newcommand{\eea}{\end{eqnarray}}
\newcommand{\bit}{\begin{itemize}}
\newcommand{\eit}{\end{itemize}}

\begin{document}

\title{Note on the symplectic structure of asymptotically flat gravity and BMS symmetries}
\author{Francesco Alessio}
\email{falessio@na.infn.it}
\affiliation{Dipartimento di Fisica ``E. Pancini" and INFN, Universit\`a degli studi di Napoli ``Federico II", I-80125 Napoli, Italy \\}

\author{Michele Arzano}
\email{michele.arzano@na.infn.it}
\affiliation{Dipartimento di Fisica ``E. Pancini" and INFN, Universit\`a degli studi di Napoli ``Federico II", I-80125 Napoli, Italy\\}

\maketitle
\begin{center}
\textbf{Abstract}
\end{center}
The Poisson brackets of the gravitational field at null infinity play a pivotal role in establishing the equivalence between the Ward identities involving Bondi-Metzner-Sachs (BMS) charges and the soft graviton theorem. In recent literature it was noticed that, in order to reproduce the action of BMS transformations via such Poisson brackets, one needs to add \textit{ad hoc} boundary terms in the symplectic form. In this article we show that, introducing a suitable splitting of the gravitational field in bulk and boundary degrees of freedom and using techniques of covariant phase space formalism, it is possible to obtain the correct Poisson brackets between the boundary fields without any additional assumption. The same Poisson brackets are used to show that BMS charges canonically generate BMS transformations on the gravitational phase space.

\section{Introduction}
The renewed interest in the infrared structure of asymptotically flat gravity has led to the discovery \cite{Strominger:2013jfa,He:2014laa} of a connection between asymptotic symmetries, in particular of Bondi-Metzner-Sachs (BMS) symmetry \cite{Sachs:1962zza}, and Weinberg's soft graviton theorem \cite{Weinberg:1965nx,Weinberg:1995mt}. These apparently uncorrelated subjects were shown to be two sides of one coin: the quantum Ward identities associated to the supertranslation and superrotation symmetry of the gravitational $\mathcal{S}$-matrix are equivalent to the leading and subleading orders of the soft graviton theorem, respectively  \cite{Strominger:2013jfa,He:2014laa,Cachazo:2014fwa,Kapec:2014opa,Strominger:2017zoo,Campiglia:2014yka,Campiglia:2015yka,Compere:2018ylh}. Moreover, these subjects were shown to be just two of the three corners of a triangular equivalence relation, the third corner consiting of the gravitational memory effect  \cite{Strominger:2014pwa,Pasterski:2015tva,Compere:2016hzt}. One of the attractive features of such infrared triangle relies in its universal character. Indeed, a similar infrared behaviour is shared by other gauge theories, including electromagnetism and strong interactions  \cite{Strominger:2013lka,Lysov:2014csa,He:2014cra}.

The description of the symplectic structure of the gravitational field at null infinity $\mathscr{I}$ for asymptotically flat spacetimes and of the associated BMS charges was originally carried out in a series of works by Ashtekar and collaborators \cite{Ashtekar:1981sf,Ashtekar:1981bq,Ashtekar:1981hw,Ashtekar:1987tt}. In the recent literature on the connection between BMS symmetries and soft graviton theorems, it was noticed however that in order to reproduce the action of supertranslations on the class of spacetimes under consideration, it is necessary to consider additional Poisson brackets involving boundary degrees of freedom of the gravitational field, which are closely related to the existence of soft gravitons \cite{Strominger:2013jfa,He:2014laa} . It has been argued, see e.g. \cite{Campiglia:2014yka,Mohd:2014oja} that such Poisson brackets can be obtained by suitably adding ``ad-hoc" boundary terms to the symplectic form, leading to boundary contributions to the Poisson brackets. In this note we show that, by decomposing the free gravitational data into a bulk and a boundary field, the symplectic form, derived using the tools of covariant phase space formalism \cite{Wald:1999wa,Compere:2018aar}, naturally provides the correct bulk-bulk and boundary-boundary Poisson brackets. In addition, we construct both the supertranslations and superrotations charges and show that they can be decomposed into a bulk and a boundary term and that they canonically generate, through the Poisson brackets defined above, BMS transformation on the phase space of the asymptotic gravitational field.\\

We start in the next Section with a brief review of the structure of gravity at null infinity, the notion of asymptotic flatness and of BMS symmetries and show how the latter act on the free gravitational data on future null infinity $\mathscr{I}^+$. In Section \ref{SEC3} we explicitly calculate the symplectic form for the gravitational field on $\mathscr{I}^+$ and in Section \ref{SEC4}, after having decomposed the fields in a bulk and a pure boundary part, we use the symplectic form to extract their Poisson brackets. In Section \ref{SEC5} we derive the supertranslation charges and prove that they canonically generate supertranslations of the fields. Finally, in Section \ref{SEC6}, we show how, after taking into account certain subtleties, our analysis can be extended to the full BMS algebra, including superrotations. We conclude with a brief summary and an outlook of the possible applications of the results presented.

\section{Asymptotic symmetries}
\label{SEC2}
In general relativity the line element of asymptotically flat spacetimes admits the following asymptotic expansion around future null infinity\footnote{A similar analysis holds for $\mathscr{I}^-$.} $\mathscr{I}^+$ \cite{Bondi:1962px,Sachs:1962wk,Tamburino:1966zz,Barnich:2010eb} in retarded Bondi coordinates\footnote{$(z,\bar{z})$ are complex stereographic coordinates on the two-sphere $S^2$: $z=\cot(\frac{\theta}{2})e^{i\varphi}$.} $(u,r,z,\bar{z})$
\begin{align}
\label{1}
\nonumber ds^2=&-du^2-2dudr+2r^2\gamma_{z\bar{z}}dzd\bar{z}\\&+\frac{2m_B}{r}du^2+rC_{zz}dz^2+rC_{\bar{z}\bar{z}}d\bar{z}^2+2g_{uz}dudz+2g_{u\bar{z}}dud\bar{z}+...
\end{align}
where the first line is just Minkowski line element, $\gamma_{z\bar{z}}$ is the metric on the unit two-sphere $S^2$ and
\begin{align}
\label{2}
g_{uz}=\frac{1}{2}D^zC_{zz}+\frac{2}{3r}N_z+\frac{2}{3r}uD_zm_B-\frac{1}{8r}D_z(C_{zz}C^{zz})+\mathcal{O}(r^{-2}),
\end{align}
with $D_z$ the covariant derivative on the two-sphere $S^2$, $m_B(u,z,\bar{z})$ the Bondi mass aspect and $N_z(u,z,\bar{z})$ the angular momentum aspect\footnote{For $N_z$ we are using the conventions of \cite{Strominger:2017zoo,Hawking:2016sgy}.}. Introducing the Bondi news function 
\begin{align}
\label{3}
N_{zz}=\partial_uC_{zz},
\end{align}
the time evolution of $m_B$ and $N_z$ is governed by Einstein equations, that for simplicity we are assuming without matter:
\begin{align}
\label{4}
&\partial_{u}m_{B}=\frac{1}{4}\left[D^2_zN^{zz}+D^2_{\bar{z}}N^{\bar{z}\bar{z}}\right]-\frac{1}{4}N_{zz}N^{zz},\\\label{4.1} &\partial_uN_z=\frac{1}{4}D_z\left[D^2_zC^{zz}-D^2_{\bar{z}}C^{\bar{z}\bar{z}}\right]-uD_z\partial_um_B+\frac{1}{4}D_z(C_{zz}N^{zz})+\frac{1}{2}C_{zz}D_zN^{zz}.
\end{align}
At first and second subleading order in the $r$ expansion the function $C_{zz}$ is the only ``free data" that we need to assign, since all the other components of the metric are determined by $C_{zz}$ through \eqref{4} and \eqref{4.1}, once initial conditions are assigned.\\

The asymptotic symmetries of asymptotically flat spacetimes were originally defined in \cite{Sachs:1962zza} as the set of diffeomorphisms that preserve the Bondi gauge and the asymptotic behaviour of the line element in \eqref{1}. They are thus generated by a vector field $\xi$ solving the following equations\footnote{In \eqref{5} and \eqref{5.5} latin indices label angular coordinates.}:
\begin{align}
\label{5}
&\mathcal{L}_{\xi}g_{rr}=0,\hspace{1cm}\mathcal{L}_{\xi}g_{rA}=0,\hspace{1cm}g^{AB}\mathcal{L}_{\xi}g_{AB}=0,\\\nonumber\\\label{5.5}
\mathcal{L}_{\xi}g_{ur}=\mathcal{O}(&r^{-2}),\hspace{1cm}\mathcal{L}_{\xi}g_{uA}=\mathcal{O}(1),\hspace{1cm}\mathcal{L}_{\xi}g_{AB}=\mathcal{O}(r),\hspace{1cm}\mathcal{L}_{\xi}g_{uu}=\mathcal{O}(r^{-1}).
\end{align}
The vector field on $\mathscr{I}^+$ satisfying these conditions is \cite{Alessio:2017lps}
\begin{align}
\label{6}
\xi|_{\mathscr{I}^+}=\left[f+\frac{u}{2}D_AY^A\right]\frac{\partial}{\partial u}+Y^A\frac{\partial}{\partial x^A},
\end{align}
where $f$ is an arbitrary function and $Y^A$ are conformal Killing vectors on $S^2$. In $(z,\bar{z})$ coordinates, this implies that $Y^z$ is holomorphic and $Y^{\bar{z}}$ is antiholomorphic. The vector field in \eqref{6} is the generator of the Bondi-Metzner-Sachs (BMS) algebra. Depending on the choice of $Y^z$ and $Y^{\bar{z}}$ we have two possible variants of such algebra. When these functions are given by $-z^{n+1}$ and $-\bar{z}^{n+1}$ with $n=-1,0,1$ we talk about ``global" BMS algebra. If $n$ can take any integer value we talk about ``local" or ``extended" BMS algebra. The former choice was the one originally implemented by Sachs \cite{Sachs:1962zza}, and the BMS algebra was defined as the semidirect sum of the algebra $\mathfrak{sl}(2,\mathbb{C})$ generating Lorentz transformations with the abelian ideal $\mathfrak{s}$ of supertranslations, consisting of arbitrary smooth functions on $S^2$. The latter choice, first suggested in \cite{Barnich:2009se,Barnich:2010eb,Barnich:2011mi} leads to an  algebra consisting of the semidirect sum of two copies of the Virasoro algebra \textbf{$\mathrm{Vir}$} (the so-called ``superrotations") with the algebra of supertranslations $\mathfrak{s}$.\\

Let us now recall how the BMS algebra acts on the free data $C_{zz}$ defining the line element in \eqref{1}. In order to obtain such action, one needs to compute the Lie derivative of the metric on-shell. The action of supertranslations on $C_{zz}$ is given by \cite{Barnich:2010eb}
\begin{align}
\label{7}
&\delta_{f} C_{zz}=fN_{zz}-2D^2_zf,\\ \label{8}&\delta_{f}N_{zz}=f\partial_uN_{zz}.
\end{align}
When $f$ reproduces an ordinary four-translation, the homogeneous term in \eqref{7} vanishes. Indeed one has $D^2_z f=D^2_{\bar{z}}f=0$ when $f$ is identified with a spherical harmonic $Y_{lm}$ with $l=0,1$. However, this is not the case for pure supertranslations, i.e. supertranslations that are not ordinary four-translations. If we start with $C_{zz}=0$, after a pure supertranslation $C'_{zz}\neq 0$ due to the homogeneous term in \eqref{7}. This has been interpreted as the fact that pure supertranslations break the vacuum of Minkowski spacetime, as discussed in \cite{Strominger:2017zoo,Strominger:2013jfa,He:2014laa,Compere:2018ylh,Compere:2016jwb}. For the action of superrotations we have
\begin{align}
\label{9}
& \delta_YC_{zz}=\frac{u}{2}D\cdot YN_{zz}+\mathcal{L}_{Y}C_{zz}-\frac{1}{2}D\cdot YC_{zz}-uD^3_zY^z,\\&
\label{9.1}
\delta_YN_{zz}=\frac{u}{2}D\cdot Y\partial_uN_{zz}+\mathcal{L}_{Y}N_{zz}-D^3_zY^z,
\end{align}
where the Lie derivative acts as
\begin{align}
\label{9.2}
\mathcal{L}_YC_{zz}=Y\cdot D C_{zz}+2D_{z}Y^zC_{zz},
\end{align} 
and similarly on $N_{zz}$. Note that the homogeneous terms in \eqref{9} and \eqref{9.1} play a similar role to those of \eqref{7} and \eqref{8}. It vanishes for $\mathfrak{sl}(2,\mathbb{C})$ transformations whereas it does not for pure superrotations. 

\section{The symplectic form on $\mathscr{I}^+$}
\label{SEC3}
In this Section we briefly review the {\it covariant phase space} approach for a generally covariant theory. Using this method we calculate, in the explicit case of general relativity, the symplectic form at future null infinity for asymptotically flat spacetimes.

Let us assume that the dynamics of the system is governed by a Lagrangian $L$. Since we are interested in the case of general relativity, we suppose that $L$ can depend both on the metric $g_{ab}$, the matter fields $\psi$ and a finite number of their derivatives. We use the collective variable $\phi\equiv(g_{ab},\psi)$ and write the change of the Lagrangian for an arbitrary variation of the fields $\phi\rightarrow\phi+\delta\phi$ as
\begin{align}
\label{10}
\delta L=\delta\phi^i\frac{\partial L}{\partial \phi^i}+\partial_{\mu}\delta\phi^i\frac{\partial L}{\partial(\partial_{\mu}\phi^i)}+...\equiv\delta\phi^i\frac{\delta L}{\delta\phi^i}+\partial_{\mu}\theta^{\mu}[\phi,\delta\phi],
\end{align}
where we have defined
\begin{align}
\label{11}
\frac{\delta L}{\delta\phi^i}\equiv\frac{\partial L}{\partial \phi^i}-\partial_{\mu}\left(\frac{\partial L}{\partial(\partial_{\mu}\phi^i)}\right)+\partial_{\mu}\partial_{\nu}\left(\frac{\partial L}{\partial(\partial_{\mu}\partial_{\nu}\phi^i)}\right)+...
\end{align}
to be the Euler-Lagrange equations of motion and the vector $\theta^{\mu}$, called ``symplectic potential", comprises all the terms that come from using repeatedly the Leibniz rule. In the language of forms \footnote{see, e.g. \cite{Compere:2018aar}.} equation \eqref{10} can be expressed as 
\begin{align}
\label{12}
\delta\mathbf{L}=\delta\phi^i\frac{\delta\mathbf{L}}{\delta\phi^i}+d\boldsymbol{\theta}[\phi,\delta\phi],
\end{align}
where $\mathbf{L}$ is the Lagrangian $4$-form and $\boldsymbol{\theta}$ the $3$-form associated with the symplectic potential. One defines the ``presymplectic current" $3$-form $\boldsymbol{\omega}$ associated to two field variations $\delta_1\phi$ and $\delta_2\phi$ as follows:
\begin{align}
\label{13}
\boldsymbol{\omega}[\phi,\delta_1\phi,\delta_2\phi]\equiv \delta_1\boldsymbol{\theta}[\phi,\delta_2\phi]-\delta_2\boldsymbol{\theta}[\phi,\delta_2\phi],
\end{align}
and, given a Cauchy surface $\Sigma$, the ``symplectic form" associated with $\Sigma$ as
\begin{align}
\label{14}
\mathbf{\Omega}_{\Sigma}[\phi,\delta_1\phi,\delta_2\phi]\equiv\int_{\Sigma}\boldsymbol{\omega}[\phi,\delta_1\phi,\delta_2\phi].
\end{align}
Note however that $\mathbf{\Omega}_{\Sigma}$ is not uniquely defined, since we have the freedom of transforming $\boldsymbol{\theta}$ as $\boldsymbol{\theta}[\phi,\delta\phi]\rightarrow\boldsymbol{\theta}[\phi,\delta\phi]+d\mathbf{Y}[\phi,\delta\phi]$ for some $2$-form $\mathbf{Y}[\phi,\delta\phi]$ which leaves \eqref{12} invariant. For details, see e.g. \cite{Wald:1999wa}.
In general, the symplectic form $\mathbf{\Omega}_{\Sigma}$ may depend on the particular slice $\Sigma$. However, if $\phi$ and the field variations $\delta_1\phi$ and $\delta_2\phi$ obey the equations of motion and the linearised equations of motion around $\phi$, respectively, provided that the integral converges, $\mathbf{\Omega}_{\Sigma}$ does not depend on the choice of $\Sigma$ \cite{Wald:1999wa}. The symplectic form is the key ingredient to define both the Poisson brackets and the conserved quantities in the theory.\\  

Here we focus on the case of General Relativity. Starting from the Einstein-Hilbert Lagrangian\footnote{$(d^{n-r}x)_{\mu_1...\mu_r}=\frac{1}{n!(n-r)!}\epsilon_{\mu_1...\mu_r\mu_{r+1}...{\mu_n}}dx^{\mu_1}\wedge...\wedge dx^{\mu_r}$ in $n$ dimensions.},
\begin{align}
\label{15}
\mathbf{L}=\frac{1}{16\pi G}\sqrt{-g}g^{\mu\nu}R_{\mu\nu}(d^4x),
\end{align}
we obtain that, under the variation $g_{\mu\nu}\rightarrow g_{\mu\nu}+h_{\mu\nu}$, i.e. $\delta g_{\mu\nu}=h_{\mu\nu}$
\begin{align}
\label{16}
\delta \mathbf{L}=-\frac{\sqrt{-g}}{16\pi G} G_{\mu\nu} h^{\mu\nu}(d^4x)+d\boldsymbol{\theta}[g,h],
\end{align}
where we have used the metric $g_{\mu\nu}$ to raise and lower the indices, $\delta g^{\mu\nu}=-g^{\mu\alpha}g^{\nu\beta}\delta g_{\alpha\beta}\equiv-h^{\mu\nu}$ and $h\equiv g^{\mu\nu}h_{\mu\nu}$ and where the symplectic potential is given by
\begin{align}
\label{19}
\boldsymbol{\theta}[g,h]=\frac{\sqrt{-g}}{16\pi G}(\nabla_{\nu}h^{\nu\mu}-\nabla^{\mu}h)(d^{3}x)_{\mu}.
\end{align}
Using \eqref{14} and \eqref{15}, the symplectic form is given \cite{Crnkovic:1986ex,Ashtekar:1990gc} by  
\begin{align}
\label{20}
\nonumber\mathbf{\Omega}_{\Sigma}[g,h_{_1},h_{_2}]=\hspace{0.1cm}\frac{1}{16\pi G}\int_{\Sigma}\sqrt{-g}\left[\frac{1}{2}h_{_2}\nabla^{\mu}h_{_1}+h_{_2\nu\rho}\nabla^{\nu}h_{_1}^{\mu\rho}-\frac{1}{2}h_{_2}\nabla_{\nu}h_{_1}^{\nu\mu}\right.\\\nonumber\\\left. -\frac{1}{2}h_{_2}^{\nu\rho}\nabla^{\mu}h_{_1\nu\rho}-\frac{1}{2}h_{_2}^{\mu\rho}\nabla_{\rho}h_{_1}-(1\leftrightarrow 2)\right](d^{3}x)_{\mu}\equiv\frac{1}{16\pi G}\int_{\Sigma}\sqrt{-g}\omega^{\mu}(d^3x)_{\mu}.
\end{align}
In general, the hypersurface $\Sigma$ can be taken either to be a spacelike slice or pushed out to a null surface. Here we are interested in asymptotically flat spacetimes and, in particular, to calculate $\mathbf{\Omega}_{\mathscr{I}^+}$, defined as
\begin{align}
\mathbf{\Omega}_{\mathscr{I}^+}\equiv\lim_{\Sigma\to\mathscr{I}^+}\mathbf{\Omega}_{\Sigma}.
\end{align}
From the line element in \eqref{1} we find that the background metric is just the Minkowski metric $\eta_{\mu\nu}$ and
\begin{align}
\label{21}
&h_{uu}=\frac{2}{r}\delta m_{B}+\mathcal{O}(r^{-2}),\hspace{1.7cm}h_{zz}=r\delta C_{zz}+\mathcal{O}(r^{-1}),\\
&h_{uz}=\frac{1}{2}D^z\delta C_{zz}+\mathcal{O}(r^{-1}),\hspace{1.2cm}h_{ur}=\mathcal{O}(r^{-2}),
\end{align}
and hence 
\begin{align}
\label{22}
&h^{rr}=\frac{2}{r}\delta m_B+\mathcal{O}(r^{-2}),\hspace{1cm}h^{rz}=-\frac{1}{2r^2}D_z\delta C^{zz}+\mathcal{O}(r^{-3}),\\&h^{zz}=\frac{1}{r^3}\delta C^{zz}+\mathcal{O}(r^{-4}),\hspace{0.81cm}h^{ur}=\mathcal{O}(r^{-2}),
\end{align}
For a null hypersurface in Bondi coordinates we have that $\sqrt{-g}(d^3x)_{\mu}=r^2\delta_{\mu}^r\gamma_{z\bar{z}}du\wedge dz\wedge d\bar{z}$ and thus we need to take the $r$ component of the integrand of \eqref{20}, that reads
\begin{align}
\label{23}
\nonumber\left.\sqrt{-g}\omega^{r}\right|_{\mathscr{I}^+}&=\gamma_{z\bar{z}}\lim_{r\to\infty}r^2\bigg[\frac{1}{2r^2}\delta_{1}C_{zz}\delta_{2}N^{zz}+\frac{1}{2r^2}\delta_{1}C_{\bar{z}\bar{z}}\delta_{2}N^{\bar{z}\bar{z}}+\mathcal{O}(r^{-3})-(1\leftrightarrow2)\bigg]\\&=\frac{1}{2}\gamma_{z\bar{z}}\left[\delta_{1}C_{zz}\delta_2N^{zz}+\delta_{1}C_{\bar{z}\bar{z}}\delta_{2}N^{\bar{z}\bar{z}}-(1\leftrightarrow2)\right].
\end{align}
In terms of the $\wedge$-product the symplectic form of \eqref{20} is then given by
\begin{align}
\label{24}
\mathbf{\Omega}_{\mathscr{I}^+}=\frac{1}{32\pi G}\int_{\mathscr{I}^+}\gamma_{z\bar{z}}\, d^2z\, du\left(\delta C_{zz}\wedge\delta N^{zz}+\delta C_{\bar{z}\bar{z}}\wedge\delta N^{\bar{z}\bar{z}}\right).
\end{align}

In order to carry out an explicit calculation of the symplectic form in \eqref{24} we need to specify the boundary conditions on the field $C_{zz}$. Such conditions play a key role because they account for the soft graviton zero modes \cite{Strominger:2013jfa,He:2014laa}. We assume that $C_{zz}$ satisfies
\begin{align}
\label{25}
\lim_{u\rightarrow\infty}C_{zz}(u,z,\bar{z})=\varphi^+_{zz}(z,\bar{z}),\hspace{1cm}\lim_{u\rightarrow-\infty}C_{zz}(u,z,\bar{z})=\varphi^-_{zz}(z,\bar{z}),
\end{align}
where $\varphi^{\pm}_{zz}(z,\bar{z})$ are smooth, non vanishing functions on $S^2$ and thus, integrating \eqref{3}, we have
\begin{align}
\label{25.1}
\int_{-\infty} ^{\infty}du\, N_{zz}=C_{zz}\Big|_{-\infty}^{\infty}=\varphi^+_{zz}-\varphi^-_{zz}\equiv \Delta\varphi_{zz}.
\end{align}
This last equation can be seen as the $\lim_{\omega\to 0}N_{zz}^{\omega}$, where $N_{zz}^{\omega}$ is the Fourier transform of $N_{zz}$. A non-vanishing $\Delta\varphi_{zz}$ measuring the difference between the gauge field $C_{zz}$ at $\mathscr{I}^+_+$ and $\mathscr{I}^+_-$, future and past of $\mathscr{I}^+$, respectively, can be associated to the existence of soft gravitons (see \cite{Strominger:2013jfa,He:2014laa,Ashtekar:2018lor} for further details).

\section{Poisson brackets of bulk and boundary fields}
\label{SEC4}
In order to introduce a decomposition of $C_{zz}$ into bulk and boundary fields we integrate equation \eqref{3} taking into account the boundary conditions \eqref{25} and write
\begin{align}
\label{25.2}
&C_{zz}(u,z,\bar{z})-\varphi^-_{zz}(z,\bar{z})=\int_{-\infty}^uN_{zz}(u',z,\bar{z})du',\hspace{0.5cm}\\&\varphi^+_{zz}(z,\bar{z})-C_{zz}(u,z,\bar{z})=\int_{u}^{\infty}N_{zz}(u',z,\bar{z})du'\,,
\end{align}
which subtracted lead to the following decomposition
\begin{align}
\label{26}
C_{zz}(u,z,\bar{z})=\frac{1}{2}\Delta\varphi_{zz}(z,\bar{z})+\varphi_{zz}^-(z,\bar{z})+\hat{C}_{zz}(u,z,\bar{z}),
\end{align}
where
\begin{align}
\label{27}
\hat{C}_{zz}(u,z,\bar{z})\equiv\frac{1}{2}\left[\int_{-\infty}^udu'N_{zz}(u',z,\bar{z})-\int_{u}^{\infty}du'N_{zz}(u',z,\bar{z})\right].
\end{align}
In \eqref{26} we are choosing $\Delta\varphi_{zz}$ and $\varphi^-_{zz}$ as independent degrees of freedom, but we could have equally chosen $\varphi^+_{zz}$ and $\varphi^-_{zz}$. Our choice is motivated by the fact that $\Delta\varphi_{zz}$ and $\varphi^-_{zz}$ will be paired in the symplectic form, i.e. they will be symplectic partners. It is also important to notice that from equation \eqref{26} we have $N_{zz}=\partial_{u}C_{zz}=\partial_u\hat{C}_{zz}$ and from \eqref{27} 
\begin{align}
\label{27.1}
\lim_{u\rightarrow\infty}\hat{C}_{zz}=\frac{1}{2}\Delta\varphi_{zz},\hspace{1cm}\lim_{u\rightarrow-\infty}\hat{C}_{zz}=-\frac{1}{2}\Delta\varphi_{zz}
\end{align}
so that \eqref{25.1} is not spoiled:
\begin{align}
\label{27.2}
\int_{-\infty}^{\infty}duN_{zz}=\left.\hat{C}_{zz}\right|_{-\infty}^{\infty}=\Delta\varphi_{zz}.
\end{align}

So far we have divided the free data $C_{zz}$ into a ``bulk" contribution $\hat{C}_{zz}$ and a pure boundary part. One of the advantages of working with $\hat{C}_{zz}$ is that it simplifies the calculation of the symplectic form, for it has the property
\begin{align}
\label{27.3}
\nonumber&\int_{-\infty}^{\infty}du \delta\hat{C}_{\bar{z}\bar{z}}\wedge\delta N^{\bar{z}\bar{z}}=-\int_{-\infty}^{\infty}du\delta N_{\bar{z}\bar{z}}\wedge\delta \hat{C}^{\bar{z}\bar{z}}+\left.\delta\hat{C}_{\bar{z}\bar{z}}\wedge\delta\hat{C}^{\bar{z}\bar{z}}\right|^{\infty}_{-\infty}=-\int_{-\infty}^{\infty}du\delta N_{\bar{z}\bar{z}}\wedge\delta \hat{C}^{\bar{z}\bar{z}}\\&=\int_{-\infty}^{\infty}du\delta \hat{C}_{zz}\wedge\delta N^{zz},
\end{align}
since the boundary term in the first line vanishes identically due to \eqref{27.1} and where we have used the antisymmetry of the wedge product in the last step.\\ 

Let us now proceed to the calculation of the symplectic form. Substituting \eqref{26} in \eqref{24} we obtain
\begin{align}
\label{28}
\mathbf{\Omega}_{\mathscr{I}^+}&=\nonumber\frac{1}{32\pi G}\int_{\mathscr{I}^+}\gamma_{z\bar{z}}d^2zdu\left(\delta \hat{C}_{zz}\wedge\delta N^{zz}+\delta \hat{C}_{\bar{z}\bar{z}}\wedge\delta N^{\bar{z}\bar{z}}\right)\\\nonumber\\	\nonumber&+\frac{1}{32\pi G}\int\gamma_{z\bar{z}}d^2z	\left(\frac{1}{2}\delta\Delta\varphi_{zz}\wedge\delta\Delta\varphi^{zz}+\delta\varphi^-_{zz}\wedge\delta\Delta\varphi^{zz}\right)\\\nonumber\\	&+\frac{1}{32\pi G}\int\gamma_{z\bar{z}}d^2z	\left(\frac{1}{2}\delta\Delta\varphi_{\bar{z}\bar{z}}\wedge\delta\Delta\varphi^{\bar{z}\bar{z}}+\delta\varphi^-_{\bar{z}\bar{z}}\wedge\delta\Delta\varphi^{\bar{z}\bar{z}}\right).
\end{align}
Using \eqref{27.3} for the second term in the first line we obtain
\begin{align}
\label{29}
\mathbf{\Omega}_{\mathscr{I}^+}&=\nonumber\frac{1}{16\pi G}\int_{\mathscr{I}^+}\gamma_{z\bar{z}}d^2zdu\delta \hat{C}_{zz}\wedge\delta N^{zz}\\\nonumber\\&+\frac{1}{32\pi G}\int\gamma_{z\bar{z}}d^2z	\left(\delta\varphi^-_{zz}\wedge\delta\Delta\varphi^{zz}+\delta\varphi^-_{\bar{z}\bar{z}}\wedge\delta\Delta\varphi^{\bar{z}\bar{z}}\right).
\end{align}
where we have used the fact that the first terms in the second and third lines of \eqref{28} cancel each other because of the antisymmetry of the $\wedge$-product. We thus see that, according to the decomposition \eqref{26}, the symplectic form splits into a bulk and a boundary part. 

In order to make contact with \cite{Strominger:2017zoo,Strominger:2013jfa,He:2014laa} we now focus on the class of spacetimes considered in these works, the so-called Christoudoulou-Klainerman (C-K) spacetimes. Such space-times are characterized by a fall-off of the Bondi news $N_{zz}$ as $u^{-1-\epsilon}$, with $\epsilon >0$, for $u\rightarrow\pm\infty$, so that the integral over $u$ in \eqref{29} converges and where
\begin{align}
\label{38.1}
\varphi^-_{zz}=D^2_zC^-,\hspace{1cm}\varphi^+_{zz}=D^2_{z}C^+,\hspace{1cm}\Delta\varphi_{zz}=D^2_zN.
\end{align}
whit $C^-$ and $C^+$ arbitrary real functions on $S^2$. These properties define our phase space $\Gamma$:
\begin{align}
\label{38.125}
\Gamma:=\{C_{zz}: C_{zz}|_{\mathscr{I}^+_{\pm}}=D^2_{z}C^{\pm}+\mathcal{O}(u^{-\epsilon}), \hspace{0.16cm}\epsilon >0\}.
\end{align}
Note that, interpreting \eqref{7} as a gauge transformation, equations \eqref{38.1} are telling us that the field $C_{zz}$ is pure gauge on $\mathscr{I}^+_{\pm}$. For the calculation of the symplectic form we also take the variations $\delta C_{zz}$ and $\delta N_{zz}$ of $C_{zz}$ and $N_{zz}$ to be C-K in the sense that
\begin{align}
\label{38.15}
&\delta C_{zz}|_{\mathscr{I}^+_{\pm}}=C'^{\pm}_{zz}|_{\mathscr{I}^+_-}-C^{\pm}_{zz}|_{\mathscr{I}^+_-}=D^2_zC'^{\pm}-D^2_zC^{\pm}=D^2_z\delta C^{\pm},\\
&\delta N_{zz}\xrightarrow {u\to\pm\infty}0\hspace{0.5cm}\mathrm{as}\hspace{0.15cm}u^{-1-\epsilon},\hspace{0.5cm}\epsilon>0.
\end{align}
For notational simplicity, from now on, we set $C^-\equiv C$. 

We can finally write the symplectic form \eqref{29} on the phase space $\Gamma$ as\footnote{We use the property $D^2_zD^2_{\bar{z}}f=D^2_{\bar{z}}D^2_{z}f$ or any smooth $f$ on $S^2$.}
\begin{align}
\label{38.2}
\mathbf{\Omega}_{\mathscr{I}^+}=\frac{1}{16\pi G}\int_{\mathscr{I}^+}\gamma_{z\bar{z}}d^2zdu\delta \hat{C}_{zz}\wedge\delta N^{zz}+\frac{1}{16\pi G}\int\gamma_{z\bar{z}}d^2zD^2_z\delta C\wedge D^{2z}\delta N.
\end{align}
Clearly, $\mathbf{\Omega}_{\mathscr{I}^+}$ converges on the whole phase space $\Gamma$. From such symplectic form we can easily read off the non-vanishing Poisson brackets:
\begin{align}
\label{30}
&\{N_{\bar{z}\bar{z}}(u,z,\bar{z}),\hat{C}_{ww}(u',w,\bar{w})\}=16\pi G\delta^{2}(z-w)\delta(u-u')\gamma_{z\bar{z}},\\\nonumber\\\label{31}&\{D^2_{\bar{z}}N(z,\bar{z}),D^2_w C(w,\bar{w})\}=16\pi G\delta^2(z-w)\gamma_{z\bar{z}}.
\end{align}
These brackets match those derived in earlier works, see e.g. \cite{He:2014laa}, however in our approach it is not necessary to add ``ad hoc" boundary terms in the symplectic form to obtain the desired result, as suggested in \cite{Campiglia:2014yka,Mohd:2014oja}. In fact, as we showed, the bulk-bulk and boundary-boundary Poisson brackets are obtained directly from the definition of the symplectic form and from the splitting \eqref{26} we introduced.
\section{Supertranslation charges as canonical generators}
\label{SEC5}
In the covariant phase space approach the infinitesimal charge associated with the asympotic symmetry generated by a vector field $\xi$ is defined as \cite{Compere:2018aar}:
\begin{align}
\label{S1}
\cancel{\delta}Q_{\xi}[\phi,\delta\phi]=\int_{\mathscr{I}^+}\boldsymbol{\omega}[\delta\phi,\delta_{\xi}\phi]=\mathbf{\Omega}_{\mathscr{I}^+}[\delta\phi,\delta_{\xi}\phi],
\end{align}
where we have introduced the notation $\cancel{\delta}$ in order to emphasize that \eqref{S1} might not be an exact differential in the field space. The finite charge can be obtained by integrating $\cancel{\delta} Q_{\xi}$ along a path in the field space. Such charge is said to be integrable if the integral does not depend on the particular path chosen, i.e. if there exists a functional $Q_{\xi}$ such that $\cancel{\delta}Q_{\xi}=\delta(Q_{\xi})$. It can be shown that $Q_{\xi}$ is conserved on shell \cite{Compere:2018aar}, that in this context means when $\phi$ satisfies the ordinary equations of motion and $\delta\phi$ the linearized equations around the solution $\phi$.\\

In order to obtain the supertranslation charges we need to find the explicit expressions for $\delta_f\hat{C}_{zz}$, $\delta_f N$ and $\delta_fC$. Taking into account \eqref{38.1}, evaluating \eqref{7} on $\mathscr{I}^+_{\pm}$ yields, due to the fall-off of the Bondi news:
\begin{align}
\label{S2}
\delta_fC_{zz}|_{\mathscr{I}^+_{+}}=D^2_z\delta_f C^+=-2D^2_zf,\hspace{0.5cm}\delta_fC_{zz}|_{\mathscr{I}^+_{-}}=D^2_z\delta_f C=-2D^2_zf,
\end{align}
so that $\delta_f C^+=\delta_f C=-2f$ and thus $\delta_f N=0$. For the bulk part we simply have
\begin{align}
\label{S3}
\delta_f\hat{C}_{zz}=fN_{zz}.
\end{align}
These expressions show that the boundary conditions of $\hat{C}_{zz}$ are preserved under a supertranslation. Indeed, using the fall-off of the Bondi news we have
\begin{align}
\label{S4}
\lim_{u\to\pm\infty}\delta_f\hat{C}_{zz}=\pm\frac{1}{2}\delta_fD^2_zN=0,
\end{align}
as required by \eqref{27.1}. We thus have that the action of supertranslations is well-defined on $\Gamma$, i.e. it maps $\Gamma$ into itself.

Using equation \eqref{S1} we can write the infinitesimal supertranslation charges as
\begin{align}
\label{S5}
\nonumber \mathbf{\Omega}_{\mathscr{I}^+}[\delta\phi,\delta_{f}\phi]\equiv\hspace{0.1cm}& \cancel{\delta}Q_{f}=\frac{1}{16\pi G}\int_{\mathscr{I}^+}\gamma_{z\bar{z}}\, d^2z\, du\left[\delta \hat{C}_{zz}\delta_{f}N^{zz}-\delta_{f}\hat{C}_{zz}\delta N^{zz}\right]\\
 &+\frac{1}{16\pi G}\int_{\mathscr{I}^+}\gamma_{z\bar{z}}\, d^2z \left[D^2_z\delta CD^{2z}\delta_fN- D^2_z\delta_{f}C D^{2z}\delta N\right].
\end{align}
Plugging equations \eqref{8} and \eqref{S4} in the previous expression, integrating by parts and using the vanishing of $N_{zz}$ on the boundaries of $\mathscr{I}^+$ we have that $\cancel{\delta}Q_f=\delta Q_f$, where
\begin{align}
\label{S7}
 Q_f=\hspace{0.1cm}-\frac{1}{16\pi G}\int_{\mathscr{I}^+}\gamma_{z\bar{z}}\, d^2z\, dufN_{zz}N^{zz}+\frac{1}{8\pi G}\int \gamma_{z\bar{z}}\, d^2z\, D^2_zfD^{2z} N \equiv Q_f^{\mathcal{H}}+Q_f^{\mathcal{S}}\,.
\end{align}
The supertranslation charge $Q_f$ thus splits into a {\it hard} and {\it soft} part $Q_{f}^{\mathcal{H}}$, and $Q_{f}^{\mathcal{S}}$, quadratic and linear in the fields, respectively. Note that in the case of an ordinary four-translation the soft term vanishes whereas in the case of a pure supertranslation it does not and its contribution is proportional to the soft mode. 
Our result \eqref{S7} exactly matches the expression used in the literature, see e.g.  \cite{Strominger:2017zoo,Strominger:2013jfa,He:2014laa}.\\

Using the Poisson brackets \eqref{30} and \eqref{31}, it is straightforward to check that $Q_f$ canonically generates supertranslations:
\begin{align}
\label{S8}
\{Q_f,\hat{C}_{zz}\}=-fN_{zz},\hspace{0.8cm}\{Q_f,D^2_zN\}=0,\hspace{0.8cm}\{Q_f,D^2_zC\}=2D^2_zf,
\end{align}
so that
\begin{align}
\label{S9}
\{Q_f,C_{zz}\}=-\delta_f C_{zz},\hspace{1cm}\{Q_f,N_{zz}\}=-\delta_fN_{zz}.
\end{align}
These relations play a central role in the recently discovered connection between the asymptotic symmetries of asimptotically flat spacetimes and soft gravitons theorems. Indeed, assuming the invariance of the gravitational $\mathcal{S}$-matrix under a diagonal $\mathrm{BMS}^0$ supertranslation\footnote{The diagonal $\mathrm{BMS}^0$ is obtained from $\mathrm{BMS}^+\times\mathrm{BMS}^-$ by means of an antipodal identification identification of the generators \cite{Strominger:2013jfa}.}, it was shown in \cite{Strominger:2013jfa,He:2014laa} that the Ward identities associated to $Q_f$ given in \eqref{S7} are equivalent to Weinberg's soft graviton theorem  \cite{Weinberg:1965nx,Weinberg:1995mt}, which relates the scattering amplitude of an arbitrary quantum process involving soft gravitons to the same amplitude without the gravitons insertion. The proportionality factor between such amplitudes is, at leading order in the soft expansion, the so-called ``soft factor" and it is related to $Q_f^{\mathcal{S}}$.

\section{Superrotations}
\label{SEC6}

A natural step at this point would be to extend the construction above to the full BMS algebra, i.e. to include superrotations in the picture. This is however not as straightforward as it might seem. Indeed, as already pointed out in \cite{Strominger:2017zoo}, the actions \eqref{9} and \eqref{9.1} can map the fields outside the original phase space $\Gamma$ in \eqref{38.125} one started with. This is because the transformed field $\delta_YC_{zz}$ diverges linearly in $u$ due to the last term in \eqref{9} and $\delta_YN_{zz}$ does not fall as $u^{-1-\epsilon}$ on the boundaries of $\mathscr{I}^+$, but gets shifted instead. The action of superrotations is in fact defined on a phase space larger than $\Gamma$, see e.g. \cite{Campiglia:2014yka,Campiglia:2015yka}. Hence, there appears to be an obstruction to the introduction of a splitting of the transformed field $\delta_YC_{zz}$ similar to that of \eqref{S3} which is preserved under superrotations.

Superrotations are also peculiar for what concern their role in the connection between the soft graviton theorem and the BMS Ward identities. Indeed while the sub-leading order in the soft expansion of the soft graviton theorem \cite{Cachazo:2014fwa}, also known as ``Cachazo-Strominger" theorem, implies the Ward identities associated to the superrotation invariance of the gravitational $\mathcal{S}$-matrix \cite{Kapec:2014opa}, {\it the converse is not true}. A complete equivalence was shown in \cite{Campiglia:2014yka,Campiglia:2015yka}, where a ``generalized BMS" algebra involving $\mathrm{Diff}(S^2)$ rather than two copies of $\mathrm{Vir}$  was considered and a new set of $\mathrm{Diff}(S^2)$ charges was introduced.

In this section, in order to by-pass the issue concerning the action of superrotations on $\Gamma$, we do not impose restrictions on the phase space and, using the tools developed in Section \ref{SEC5}, we derive the infinitesimal charges associated to the transformations \eqref{9} and \eqref{9.1}. As it will turn out, they will consist of an integrable and a non-integrable part, in agreement with the result of \cite{Barnich:2011mi}. We will show how their components split into bulk and boundary terms, both of which comprise a hard and a soft component and we prove that the integrable part canonically generates the transformations of the fields, using the Poisson brackets of \eqref{30} and \eqref{31}. Our results for the bulk degrees of freedom will reproducee the expression for the charges given in in the literature in \cite{Campiglia:2014yka,Campiglia:2015yka}.

In order to obtain the infinitesimal charges, let us derive the transformation laws of the bulk and boundary fields under superrotations. From equation \eqref{9} we have
\begin{align}
\label{R5}
&D^2_z\delta_YC=\mathcal{L}_YD^2_zC-\frac{1}{2}D\cdot YD^2_zC+\lim_{u\to\infty}uD^3_zY^z,\\
\label{R5.1}
&D^2_z\delta_YN=\mathcal{L}_YD^2_zN-\frac{1}{2}D\cdot YD^2_zN-2\lim_{u\to\infty}uD^3_zY^z,\\
\label{R5.2}
&\delta_Y\hat{C}_{zz}=\frac{u}{2}D\cdot YN_{zz}+\mathcal{L}_Y\hat{C}_{zz}-\frac{1}{2}D\cdot Y\hat{C}_{zz}-uD^3_zY^{z}.
\end{align}
As remarked above, the transformed boundary terms are divergent. However, as we will see, the charges we will derive below generate exactly the transformations in \eqref{R5}, \eqref{R5.1} and \eqref{R5.2}.

The infinitesimal charge is given by $\cancel{\delta}Q_Y=\mathbf{\Omega}_{\mathscr{I}^+}[\delta\phi,\delta_Y\phi]$ and, as in the case of supertranslations, it is divided into a bulk part and a boundary part. Let us analyze them separately. The bulk part is given by
\begin{align}
\label{R6}
\nonumber\cancel{\delta}\hat{Q}_Y=&\frac{1}{16\pi G}\int_{\mathscr{I}^+}\gamma_{z\bar{z}}\, d^2z\, du\Big[\delta\hat{C}_{zz}\Big(\frac{u}{2}D\cdot Y\partial_uN^{zz}+\mathcal{L}_YN^{zz}-D^{2z}D_{\bar{z}}Y^{\bar{z}}\Big)\\&-\delta N^{zz}\Big(\frac{u}{2}D\cdot YN_{zz}+\mathcal{L}_Y\hat{C}_{zz}-\frac{1}{2}D\cdot Y\hat{C}_{zz}-uD^3_zY^z\Big)\Big]\,.
\end{align}
Integrating by parts it is easy to see that this expression decomposes in an integrable part $\delta \hat{Q}_Y$ and a non-integrable contribution $\hat{\Theta}_Y$. The charge associated to the integrable part is given by\footnote{The decomposition is non-unique, since it is invariant under the transformation $\hat{Q}'_Y=\hat{Q}_Y-F[\phi]$ and $\Theta'_Y=\Theta_Y+\delta F[\phi]$, for some functional $F[\phi]$.}
\begin{multline}\label{R7}
\hat{Q}_Y=\frac{1}{16\pi G}\int_{\mathscr{I}^+}\gamma_{z\bar{z}}\, d^2z\, du\Big[-\frac{u}{2}D\cdot YN_{zz}N^{zz}+\\
+\frac{1}{2}D\cdot Y\hat{C}_{zz}N^{zz}-N^{zz}\mathcal{L}_Y\hat{C}_{zz}-D^{2z}D_{\bar{z}}Y^{\bar{z}}\hat{C}_{zz}+uD^3_zY^{z}N^{zz}\Big],
\end{multline}
while the non integrable part is
\be
\label{R7.0}
\hat{\Theta}_Y=-\frac{1}{8\pi G}\int_{\mathscr{I}^+}\gamma_{z\bar{z}}\, d^2z\, du\, D\cdot Y\delta\hat{C}_{zz}N^{zz}\,.
\ee
Note that, as we anticipated above, integrating by parts\footnote{For the boundary term to be neglected we should impose that $\hat{C}_{zz}\xrightarrow {u\to\pm\infty} u^{-1-\epsilon}$, with $\epsilon>0$.} the last term in \eqref{R7} yields the same result obtained in \cite{Campiglia:2014yka,Campiglia:2015yka}. As for supertranslations, the charge $\hat{Q}_Y$ is composed by a hard and a soft part, $\hat{Q}_Y^{\mathcal{H}}$ and $\hat{Q}_Y^{\mathcal{S}}$, where $\hat{Q}_Y^{\mathcal{S}}$ is given by the last two terms in \eqref{R7}, linear in the fields.  It is straightforward to check, using the Poisson brackets of \eqref{30} that
\begin{align}
\label{R8}
\{\hat{Q}_Y,\hat{C}_{zz}\}=-\delta_Y\hat{C}_{zz},\hspace{1cm}\{\hat{Q}_Y,N_{zz}\}=-\delta_YN_{zz}.
\end{align}
i.e. the charge obtained from the integrable part of $\cancel{\delta}\hat{Q}_Y$ canonically generates superrotations on the bulk fields.\\

Let us now focus on the boundary fields. The infinitesimal charge is now given by
\begin{align}
\label{R9}
\nonumber\cancel{\delta}\tilde{Q}_{Y}=\frac{1}{16\pi G}\int&\gamma_{z\bar{z}}\, d^2z\, \Big[D^2_z\delta C\Big(\mathcal{L}_YD^{2z}N-\frac{1}{2}D\cdot YD^{2z}N-2\lim_{u\to\infty} uD^{2z}D_{\bar{z}}Y^{\bar{z}}\Big)\\&-D^{2z}\delta N\Big(\mathcal{L}_YD^2_zC-\frac{1}{2}D\cdot YD^2_zC+\lim_{u\to\infty}uD^3_zY^{z}\Big)\Big].
\end{align}
Similarly to the bulk part, $\cancel{\delta}\tilde{Q}_{Y}$ contains integrable a non integrable contributions leading to
\begin{align}
\label{R10}
\nonumber&\tilde{Q}_Y=\frac{1}{16\pi G}\int_{\mathscr{I}^+}\gamma_{z\bar{z}}\, d^2z\, \Big[-D^{2z}N\mathcal{L}_YD^2_zC+\frac{1}{2}D\cdot YD^{2z}ND^2_zC\\&\hspace{3cm}-2D^2_zC\lim_{u\to\infty}uD^{2z}D_{\bar{z}}Y^{\bar{z}}-D^{2z}N\lim_{u\to\infty}uD^3_zY^z\Big],\\
&\tilde{\Theta}_Y=-\frac{1}{8\pi G}\int\gamma_{z\bar{z}}\, d^2z\, D\cdot YD^2_{z}\delta CD^{2z}N.
\end{align}
The last two terms in \eqref{R10} can be collected as a soft boundary charge $\tilde{Q}_Y^{\mathcal{S}}$ while the others give the hard boundary charge $\tilde{Q}_Y^{\mathcal{H}}$. Using \eqref{31}, it can be checked that 
\begin{align}
\label{R11}
\{\tilde{Q}_Y,D^2_zC\}=-D^2_z\delta_YC.
\end{align}
With some simple algebra it is easy to show that $\tilde{Q}_Y$ of \eqref{R10} can also be written as
\begin{align}
\label{R12}
\nonumber&\tilde{Q}_Y=\frac{1}{16\pi G}\int_{\mathscr{I}^+}\gamma_{z\bar{z}}d^2z\Big[D^{2\bar{z}}C\mathcal{L}_YD^2_{\bar{z}}N-\frac{1}{2}D\cdot YD^{2\bar{z}}CD^2_{\bar{z}}N\\&\hspace{3cm}-2D^2_zC\lim_{u\to\infty}uD^{2z}D_{\bar{z}}Y^{\bar{z}}-D^{2z}N\lim_{u\to\infty}uD^3_zY^z\Big],
\end{align}
from which it is easy to see that
\begin{align}
\label{R13}
\{\tilde{Q}_Y,D^2_zN\}=-D^2_z\delta_YN.
\end{align}
Putting \eqref{R8}, \eqref{R11} and \eqref{R13} together, on defining $Q_Y=\hat{Q}_Y+\tilde{Q}_Y$, we see that
\begin{align}
\label{R14}
\{Q_Y,C_{zz}\}=-\delta_YC_{zz},
\end{align}
i.e. the charge $Q_Y$ canonically generates superrotations.

\section{Conclusions}

In this note we provided new insights on the phase space structure of asymptotically flat gravity and its asymptotic symmetries. Using the tools of covariant phase space formalism we first derived the symplectic form of general relativity at null infinity in the asymptotically flat regime. We then introduced a decomposition of the gravitational free data on $\mathscr{I}^+$ into bulk and boundary contributions. We showed how, under such decomposition, the symplectic form undergoes a similar splitting in bulk and boundary terms suggesting that bulk and boundary fields decouple completely from each other and can be thus treated as independent degrees of freedom. Through such decomposition we were able to reproduce the Poisson brackets for asymptotically flat gravity known in the literature without the unpleasant drawback of having to introduce ``by hand" additional boundary contributions. 

The last part of this note was devoted to the application of the tools developed to derive the BMS conserved charges of the theory and show that via their Poisson brackets with the fields they canonically generate BMS transformations. For supertranslations this task was straightforward with the conserved charge decomposing in the well known hard and soft contributions. For superrotations the derivation of the conserved charges was slightly more involved since the variation of the symplectic form produced non-integrable contributions which had to be discarded while the integrable terms led to conserved charges, again splitting in hard and soft contributions, which canonically generate superrotations.

The decoupling in bulk and boundary degrees of freedom at null infinity we uncovered in this work could be relevant in several contexts. On one side this bulk/boundary decomposition is particularly suggestive since the boundary conditions on the field $C_{zz}$ play a primary role in establishing the equivalence between the BMS Ward identities and Weinberg's soft graviton theorem. Thus it is quite natural to ask which insights could our phase space construction provide in a quantum setting, particularly for what concerns the role of the zero modes of the gravitational field \cite{Ashtekar:2014zsa, Ashtekar:2018lor}. On the other hand the independent action of the BMS algebra on the bulk and boundary components of the gravitational field which we spelled out might have interesting applications for what concerns the ``holographic" aspects of the description of flat space scattering amplitudes as correlators on the celestial sphere \cite{Kapec:2016jld, Cheung:2016iub,Pasterski:2016qvg}. This could have useful ramifications for the ambitious programme of setting up an holographic description of four dimensional quantum gravity in terms of a conformal field theory living on the celestial sphere \cite{Ball:2019atb}. We postpone to future studies a more in depth exploration of these matters.

\end{document}